\documentclass[%
reprint,
 amsmath,amssymb,
 aps,pra
]{revtex4-1}

\usepackage{graphicx}% Include figure files
\usepackage{dcolumn}% Align table columns on decimal point
\usepackage{bm}% bold math
\usepackage[colorlinks=true, allcolors=blue]{hyperref}
\begin{document}

\preprint{APS/123-QED}

\title{Non-adiabatic Effects Induced by Strong Light-Matter Coupling in Cavity QED}
\author{T. Zalialiutdinov$^{1,2}$}
\email[E-mail:]{t.zalialiutdinov@spbu.ru}
\author{D. Solovyev$^{1,2}$}
\author{J. J. Lopez-Rodriguez$^{1}$}
\author{A. Anikin$^{1,3}$} 
\author{A. Kotov$^{1}$}
\affiliation{$^1$Department of Physics, St. Petersburg State University, Petrodvorets, Oulianovskaya 1, 198504, St. Petersburg, Russia}
\affiliation{$^2$Petersburg Nuclear Physics Institute named after B.P. Konstantinov of National Research Centre 'Kurchatov Institut', St. Petersburg, Gatchina, 188300, Russia}
\affiliation{$^3$D. I. Mendeleev Institute for Metrology, St. Petersburg, 190005, Russia}

\date{\today}% It is always \today, today,

\begin{abstract}
We present a systematic study of the diagonal Born–Oppenheimer correction (DBOC) for atoms and molecules embedded in optical cavities and interacting with a quantized electromagnetic field. By explicitly evaluating the nuclear kinetic energy operator, we analyze cavity-induced modifications of DBOC within a quantum electrodynamics configuration–interaction (QED-CI) framework built on quantum electrodynamics Hartree–Fock (QED-HF) and strong-coupling quantum electrodynamics Hartree–Fock (SC-QED-HF) reference states. The analysis covers a diverse set atomic and molecular systems, including He, H$^{-}$, Be, H$_2$, LiH, HF, ammonia (NH$_3$), and formaldehyde (CH$_2$O). We show that the presence of the cavity leads to shifts in molecular dissociation energies on the order of a few inverse centimeters. For several atomic systems, the inclusion of the DBOC yields a pronounced effect, with the correction magnitude reaching the experimental resolution. These findings reveal finite nuclear mass effects as an essential component of nuclear dynamics in cavity QED and suggest their relevance for precision analysis in strongly coupled light–matter systems.
\end{abstract}

\maketitle

\section{Introduction}

The interaction of molecules with quantized electromagnetic fields
confined in optical cavities has emerged as a powerful means of
modifying molecular properties in the absence of external driving
fields~\cite{Hutchison2012,Ebbesen2016,GarciaVidal2021}. In the strong-coupling regime, the coherent hybridization of photonic and matter degrees of freedom gives rise to polariton states that persist even in the electronic ground state, resulting in measurable changes in molecular structure, chemical reaction energetics, and dynamics \cite{Stranius2018,Schlawin2022}. Such cavity-mediated effects have been observed in a broad range of physical platforms, spanning both vibrational and electronic strong-coupling regimes. They include profound modifications of chemical reactivity and substantial reshaping of molecular energy landscapes ~\cite{Thomas2016,Thomas2019,Ahn2023,Nagarajan2021,Feist2018}.
These experimental advances have motivated the development of rigorous
first-principles theoretical frameworks for molecules coupled to
quantized radiation fields~\cite{Flick2017,Mandal2023,Li2022}.

Significant progress has been made in {\it ab initio} electronic-structure methods for molecular cavity quantum electrodynamics \cite{Ruggenthaler2023}. Methods based on the Pauli-Fierz Hamiltonian, such as quantum-electrodynamics Hartree–Fock (QED-HF) \cite{Riso2022}, QED density functional theory \cite{Ruggenthaler2014,Ruggenthaler2018NRC,Flick2018QEDFT}, and QED coupled-cluster theory \cite{Haugland2020}, treat electrons and photons on equal footing. These approaches show that the cavity vacuum field can cause non-perturbative changes to electronic ground states, molecular orbitals, and potential energy surfaces, even far from resonance \cite{Riso2022,Flick2017,Ruggenthaler2023}. In particular, QED-HF calculations in the strong-coupling regime provide clear insights into how the cavity modifies the electronic structure, as well as the resulting changes in molecular geometry and response properties \cite{Riso2022}.

Cavity QED calculations typically rely on the Born-Oppenheimer (BO) approximation for many-particle systems, in which nuclear dynamics is taken into account only indirectly, for example, through nuclear vibrations and rotations or via a parametric dependence of the electronic potential energy surfaces on the internuclear distance \cite{Ruggenthaler2023, Mandal2023}. Non-adiabatic and beyond–Born–Oppenheimer effects, and in particular the diagonal Born–Oppenheimer correction (DBOC), have so far received little attention in cavity settings.

In free space, the DBOC originates from the action of the nuclear kinetic energy operator on the electronic wave function and is known to be essential for accurate predictions of molecular energetics, dissociation energies, and high-resolution spectroscopy \cite{Flick2017,Ruggenthaler2018NRC}. Since cavity fields can substantially, and often non-perturbatively, alter electronic wave functions \cite{Riso2022, Haugland2020}, both a modification via the DBOC and a modification of the DBOC itself (in its zeroth-order, i.e., cavity-free, approximation) are physically well motivated. Nevertheless, a systematic ab initio investigation of this effect has been lacking.

In this work, we present an explicit analysis of the diagonal Born–Oppenheimer correction for molecules interacting with a quantized radiation field inside an optical cavity. The DBOC is evaluated fully numerically using a quantum electrodynamics Hartree–Fock (QED-HF) reference state, augmented with single and double excitation configuration interaction (CISD) to account for electron–electron correlations. This framework is applied to both standard QED-HF calculations and those performed in the strong light–matter coupling regime, denoted as SC-QED-HF \cite{Foley2023}. We investigate a representative set of single atoms and molecules, including He, H$^{-}$, Be, H$_2$, LiH, HF, NH$_3$, and CH$_2$O, to provide a comparative assessment of cavity-induced effects across varying molecular sizes, masses, and bonding topologies.

Our results demonstrate that cavity-induced modifications to the DBOC result in systematic shifts in molecular dissociation energies on the order of few cm$^{-1}$. Far from being a mere background perturbation, these shifts are spectroscopically highly significant; for light atoms and molecules, they fall well within the resolving power of modern high-resolution experiments. These results show a previously overlooked contribution to molecular energies in cavity QED and indicate that adiabatic corrections to the Born–Oppenheimer approximation are important for accurate predictions under strong light–matter coupling.

The paper is organized as follows. Section~\ref{theory} introduces the theoretical framework of molecular cavity QED and summarizes the QED-HF approach. In Section~\ref{dboc_theory}, we detail the numerical evaluation of the DBOC in the presence of a quantized radiation field. Sections~\ref{results_atoms} and \ref{results_molecules} provide computational details and describe how adiabatic corrections to the Born–Oppenheimer approximation are treated for atoms and molecules, respectively, including a comparison between the standard QED-HF and SC-QED-HF regimes. The polyatomic molecules NH$_3$ and CH$_2$O are also studied in Section~\ref{poly}. Finally, Section~\ref{conclusion} summarizes our primary conclusions and discusses the implications for precision molecular spectroscopy and cavity-modified nuclear dynamics. Atomic units ($\hbar = m_e = e = 4\pi\epsilon_0 = 1$) are employed throughout the paper.

\section{Theoretical foundations}
\label{theory}

Below, molecular systems coupled to quantized electromagnetic fields confined within optical cavities are investigated using the framework of non-relativistic cavity quantum electrodynamics (cavity QED).
The theoretical foundation rests on the Pauli-Fierz Hamiltonian expressed in the length gauge~\cite{Ruggenthaler2018}, formulated within both the dipole approximation and the Born-Oppenheimer approximation. The dipole approximation is valid when the cavity wavelength substantially exceeds molecular dimensions ($\lambda_{\mathrm{cav}} \gg$ molecular size), permitting treatment of the electromagnetic field as spatially uniform over the molecular extent. The Born-Oppenheimer approximation separates electronic and nuclear motion by treating nuclear coordinates $\{\mathbf{R}_A\}$ as fixed parameters in the clamped-nucleus electronic Hamiltonian, with nuclear dynamics governed by a subsequent quantum-mechanical equation in which the adiabatic potential energy surface $E(\{\mathbf{R}_A\})$ acts as the effective potential. Within this framework, the electron-photon system is described by a
Hamiltonian that provides an exact, non-perturbative account of light-matter interaction derived from first principles of quantum electrodynamics, without recourse to perturbation theory or model assumptions about the molecular electronic structure.

In this study, we consider coupling to a single cavity mode with frequency $\omega_{\mathrm{cav}}$ and polarization vector $\bm{\epsilon}$. Two particular cases are of interest: alignment of $\bm{\epsilon}$ along a molecular axis and perpendicular to it. Within the dipole approximation, the many-electron system embedded in such an environment is described by the Pauli-Fierz Hamiltonian:
\begin{align}
\hat{H}_{\mathrm{PF}}
  &= \hat{H}_{\mathrm{e}}
   + \omega_{\mathrm{cav}} \hat{b}^\dagger \hat{b} \nonumber \\
  &\quad
   - \sqrt{\frac{\omega_{\mathrm{cav}}}{2}}\,
     (\bm{\lambda}\cdot \hat{\bm{d}})\,
     (\hat{b}^\dagger + \hat{b})
   + \frac{1}{2}
     (\bm{\lambda}\cdot \hat{\bm{d}})^2.
\label{eq:PF}
\end{align}
Here $\hat{H}_{\mathrm{e}}$ denotes the standard non-relativistic electronic Hamiltonian in the Born-Oppenheimer approximation:
\begin{eqnarray}
\hat{H}_{\mathrm{e}}
=
\sum_{i=1}^{N_e}
\left(
-\frac{1}{2}\nabla_i^2
-
\sum_{A=1}^{N_n}
\frac{Z_A}{|\mathbf{r}_i - \mathbf{R}_A|}
\right)
+
\sum_{i<j}^{N_e}
\frac{1}{|\mathbf{r}_i - \mathbf{r}_j|}
\\
\nonumber
+
\sum_{A<B}^{N_n}
\frac{Z_A Z_B}{|\mathbf{R}_A - \mathbf{R}_B|},
\label{eq:He}
\end{eqnarray}
comprising electronic kinetic energy, electron-nuclear attraction, interelectronic repulsion, and internuclear repulsion respectively. 
In the second and third terms of Eq. (\ref{eq:PF}), $\hat{b}^\dagger$ and $\hat{b}$ are bosonic photon creation and annihilation operators satisfying the canonical commutation relation $[\hat{b},\hat{b}^\dagger]=1$, and $\hat{\bm{d}} = -\sum_i \mathbf{r}_i + \sum_A Z_A \mathbf{R}_A$ is the total molecular dipole operator, with both electronic ($-\sum_i \mathbf{r}_i$) and nuclear ($\sum_A Z_A \mathbf{R}_A$) contributions. The coupling vector $\bm{\lambda} = \sqrt{\omega_{\mathrm{cav}}/(2\epsilon_0 V)}\,\bm{\epsilon}$ encodes the cavity mode volume $V$ and polarization direction $\bm{\epsilon}$, with $\epsilon_0$ denoting the vacuum permittivity. For Fabry-P\'erot cavities with mirror separation $L$, we have $V \approx L^3$, leading to coupling strengths $\lambda \sim L^{-3/2}$.

Three remaining terms in Eq.~(\ref{1}) warrant detailed discussion. The second term, $\omega_{\mathrm{cav}} \hat{b}^\dagger \hat{b}$, represents the free photonic field as a quantum harmonic oscillator with frequency $\omega_{\mathrm{cav}}$. The operator $\hat{b}^\dagger \hat{b}$ is the photon number operator; its eigenstates $|n\rangle$ satisfy $\hat{b}^\dagger \hat{b}|n\rangle = n|n\rangle$, where the integer eigenvalue $n = 0, 1, 2, \ldots$ gives the number of photons in the cavity mode. These Fock states form a complete orthonormal basis: $\langle n | m \rangle = \delta_{nm}$ and $\sum_{n=0}^\infty |n\rangle\langle n| = \mathbb{I}_{\mathrm{photon}}$. The creation and annihilation operators act on these states according to $\hat{b}^\dagger|n\rangle = \sqrt{n+1}|n+1\rangle$ and $\hat{b}|n\rangle = \sqrt{n}|n-1\rangle$, with $\hat{b}|0\rangle = 0$ defining the photon vacuum.

The third term, $-\sqrt{\omega_{\mathrm{cav}}/2}\,(\bm{\lambda}\cdot \hat{\bm{d}})\,(\hat{b}^\dagger + \hat{b})$, is the bilinear coupling that mediates energy exchange between matter and light. This term drives transitions between photon states while simultaneously inducing electronic polarization, and vice versa. Its form arises from the minimal coupling Hamiltonian $\sum_i (\mathbf{p}_i - e\mathbf{A}(\mathbf{r}_i))^2/(2m_e)$ in the Coulomb gauge through the Power-Zienau-Woolley unitary transformation to the length gauge~\cite{PZW}. The fourth term, $\frac{1}{2}(\bm{\lambda}\cdot \hat{\bm{d}})^2$, is referred to as the dipole self-energy. It originates from the diamagnetic term $\sim\bm{A}^2$.

A fundamental {\it ab initio} treatment of Eq.~(\ref{eq:PF}) is the quantum electrodynamic Hartree-Fock theory (QED-HF)~\cite{Haugland2020}, where the ground state is approximated as a product ansatz $|\Psi_{\mathrm{QED-HF}}\rangle = |\Phi_0^{\mathrm{e}}\rangle \otimes |\Phi_0^{\mathrm{p}}\rangle$, separating electronic and photonic degrees of freedom. Here, $|\Phi_0^{\mathrm{e}}\rangle$ denotes an electronic Slater determinant built from $N_e$ molecular spin-orbitals, and $|\Phi_0^{\mathrm{p}}\rangle$ represents the photonic state. The photonic component is conventionally expanded in the Fock basis as
\begin{eqnarray}
|\Phi_0^{\mathrm{p}}\rangle = \sum_{n=0}^{N_{\max}} c_n(\hat{b}^\dagger)^{n} |0\rangle= \sum_{n=0}^{N_{\max}} c_n |n\rangle,
\label{eq:photon_expansion}
\end{eqnarray}
where the coefficients $\{c_n\}$ are variational parameters to be determined and $N_{\max}$ is a truncation parameter. Then the probability of the state with $n$-photons is given by $p_{n}=|{c_n}|^2$. The expansion (\ref{eq:photon_expansion}) allows incorporation of virtual photon excitations (fluctuations away from the vacuum state $|0\rangle$) at the mean-field level. However, convergence with respect to $N_{\max}$ is notoriously slow, often requiring $N_{\max} \sim 20$ or more photon states to achieve sub-Hartree accuracy~\cite{Foley2023}. More critically, for charged molecular systems, the Fock-state expansion leads to an unphysical dependence of the total energy on the choice of coordinate origin. Translating the molecular system in space alters the computed energy, thereby violating translational invariance. Additionally, the QED-HF energy exhibits an artificial dependence on the cavity frequency $\omega_{\mathrm{cav}}$ when the photon basis is poorly converged~\cite{Foley2023}.

These issues are largely circumvented by recasting QED-HF within a coherent-state basis. According to \cite{Foley2023} it is necessary to apply the unitary displacement transformation 
\begin{eqnarray}
\label{CS1}
\hat{U}_{\mathrm{CS}} = \exp[z(\hat{b}^\dagger - \hat{b})].
\end{eqnarray}
In Eq. (\ref{CS1}) the displacement parameter $z$ is given by
\begin{eqnarray}
z = -\frac{\bm{\lambda}\cdot \langle \hat{\bm{d}} \rangle}{\sqrt{2\omega_{\mathrm{cav}}}},
\label{eq:displacement}
\end{eqnarray}
where $\langle \hat{\bm{d}} \rangle$ is the expectation value of the dipole operator with respect to the electronic state $|\Phi_0^{\mathrm{e}}\rangle$. This transformation diagonalizes the photonic part of the Hamiltonian $\hat{H}_{\mathrm{PF}}$ and shifts the photon creation and annihilation operators according to $\hat{U}_{\mathrm{CS}}^\dagger \hat{b} \hat{U}_{\mathrm{CS}} = \hat{b} - z$ and $\hat{U}_{\mathrm{CS}}^\dagger \hat{b}^\dagger \hat{U}_{\mathrm{CS}} = \hat{b}^\dagger - z$. In the transformed frame, the photonic ground state is simply the vacuum $|0\rangle$ in the shifted mode coordinate. Then the effective Hamiltonian in coherent state representation, see \cite{Bauman2025}, is
\begin{eqnarray}
\label{eq:HCS}
\hat{H}_{\mathrm{CS}}
=
\hat{H}_{\mathrm{e}}
+\omega_{\mathrm{cav}}\hat{b}^\dagger \hat{b} - \sqrt{\frac{\omega_{\mathrm{cav}}}{2}}
[\bm{\lambda}(\hat{\bm{d}}-\langle \hat{\bm{d}}\rangle)]
\\\nonumber
\times(\hat{b}^\dagger + \hat{b})
+
\frac{1}{2}
[\bm{\lambda}\cdot (\hat{\bm{d}} - \langle \hat{\bm{d}} \rangle)]^2,
\end{eqnarray}
where cavity effects now manifest solely through fluctuations in the dipole moment $(\hat{\bm{d}} - \langle \hat{\bm{d}} \rangle)$ rather than the dipole moment itself. After transformation, the nuclear dependence in the dipole operator vanishes, leaving only the electronic contribution $\hat{\bm{d}}$. As a result, photonic degrees of freedom no longer require self-consistent optimization. 

This fluctuation-based formulation is physically transparent: the cavity field couples to deviations of the instantaneous dipole from its mean value, thereby stabilizing or destabilizing charge distributions depending on their fluctuation properties. In the coherent-state representation, where the Hamiltonian is given by Eq.~(\ref{eq:HCS}), the wave function is approximated by the ansatz $|\Psi_{\mathrm{QED-HF}}^{\mathrm{CS}}\rangle = |\Phi_0^{\mathrm{e}}\rangle \otimes |0\rangle$. As a result the photon field contributes to the hamiltonian only through the dipole self-energy (DSE) term:
\begin{eqnarray}
\label{eq:HCS2}
\hat{H}_{\mathrm{CS}}
=
\hat{H}_{\mathrm{e}}
+\frac{1}{2}
[\bm{\lambda}\cdot (\hat{\bm{d}} - \langle \hat{\bm{d}} \rangle)]^2.
\end{eqnarray}
The details of the implementation of the DSE term in the SCF procedure can be found, for example, in \cite{Vu2022EnhancedDiastereocontrol,McTague2022NonHermitianCQEDCIS,DePrince2021CavityModulatedIPEA,Mallory2022ReducedDensityMatrix,Liebenthal2022EOMElectronAttachment,Foley2023}.
From Eq. (\ref{eq:HCS2}) it is clearly seen that at the mean-field level that the energy is independent of the cavity frequency $\omega_{\mathrm{cav}}$ \cite{Foley2023}. Cavity‑frequency dependence arises solely upon including excitations in the photonic Fock space, i.e., when post‑mean‑field methods account for explicit electron‑photon correlations.

Accordingly, the coherent-state QED-HF formulation provides several important practical and theoretical advantages. First, it eliminates photon-basis truncation artifacts entirely - no parameter analogous to $N_{\max}$ appears. Second, it ensures origin-independent total energies for both neutral and charged molecular systems. Third, the energy becomes independent of $\omega_{\mathrm{cav}}$. However, a subtle deficiency persists: while the total energy is origin-independent, the molecular orbital coefficients and the Fock matrix itself retain residual origin dependence for charged species~\cite{Riso2022}. This represents a violation of full translational invariance that, while not affecting ground-state energies, complicates subsequent correlated treatments such as perturbation theory or coupled-cluster methods. In these methods, orbital‑dependent quantities enter explicitly.

To restore translational invariance and obtain cavity-consistent molecular orbitals in the strong-coupling regime, the SC-QED-HF method introduces an orbital-dependent coherent-state displacement \cite{Riso2022}. Rather than employing a global displacement parameter $z$, see Eq.~(\ref{eq:displacement}), each spin-orbital, $p\sigma$ ($p$ stands for the index of a spatial orbital, and $\sigma$ for the index of the spin state), is assigned its own variational dressing parameter $\eta_{p\sigma}$. The corresponding unitary transformation reads
\begin{eqnarray}
\hat{U}_{\mathrm{SC}}
=
\exp\!\left[
\frac{\bm{\lambda}}{\sqrt{2\omega_{\mathrm{cav}}}}
\sum_{p\sigma}
\eta_{p\sigma}
\hat{a}_{p\sigma}^\dagger \hat{a}_{p\sigma}
(\hat{b} - \hat{b}^\dagger)
\right],
\label{eq:SC_transformation}
\end{eqnarray}
where $\hat{a}_{p\sigma}^\dagger$ and $\hat{a}_{p\sigma}$ are the fermionic creation and annihilation operators for spin-orbital $p\sigma$. The occupation-number operator $\hat{a}_{p\sigma}^\dagger \hat{a}_{p\sigma}$ ensures that the displacement of the photonic mode is conditioned on the electronic configuration: the field is shifted only in those orbitals that are occupied, so that each occupied orbital is surrounded by its own effective photonic cloud \cite{Riso2022}.

The effective Hamiltonian is obtained by acting with $\hat{U}_{\mathrm{SC}}$ on the Pauli--Fierz Hamiltonian and projecting onto the photon vacuum. The result is given by
\begin{eqnarray}
\hat{H}_{\mathrm{SC}}
=
\hat{H}_{\mathrm{e}}
+
\frac{1}{2}
\left[
\bm{\lambda}\cdot
\left(
\hat{\bm{d}}
-
\sum_{p\sigma}
\eta_{p\sigma}
\hat{a}_{p\sigma}^\dagger \hat{a}_{p\sigma}
\right)
\right]^2.
\label{eq:HSC}
\end{eqnarray}
The operator in parentheses in the above formula
\begin{eqnarray}
\hat{d}_{\mathrm{ref}}
=
\sum_{p\sigma}
\eta_{p\sigma}
\hat{a}_{p\sigma}^\dagger \hat{a}_{p\sigma}
\end{eqnarray}
serves as an orbital-resolved reference dipole, replacing the single averaged dipole moment employed in simpler QED-HF formulations \cite{Riso2022}. The interaction with the cavity is then governed by the fluctuations of the physical dipole moment $\hat{\bm{d}}$ about this orbitally dressed reference: rather than subtracting a single global mean, the method subtracts the individual contribution of each orbital, thereby providing a more accurate account of the inhomogeneity of the electron-photon interaction.

Within a mean-field treatment, expanding the square in Eq.~(\ref{eq:HSC}) and factorizing the resulting density matrices leads to a modified Fock operator containing the standard electronic terms together with additional cavity-induced one- and two-electron contributions \cite{Riso2022}. The molecular orbital coefficients $\{C_{\mu p}\}$ and the dressing parameters $\{\eta_{p\sigma}\}$ are obtained from the coupled nonlinear equations
\begin{eqnarray}
\mathbf{F}(\mathbf{C}, \bm{\eta})\,\mathbf{C} = \mathbf{S}\,\mathbf{C}\,\bm{\varepsilon}.
\label{eq:gen_eig_SCQEDHF}
\end{eqnarray}
Here $\mathbf{F}$ is the cavity-modified Fock matrix, $\mathbf{S}$ is the overlap matrix, and $\bm{\varepsilon}$ is the diagonal matrix of orbital energies. The parameters $\eta_{p\sigma}$ are determined from the stationarity conditions $\partial E / \partial \eta_{p\sigma} = 0$. At each iteration of the self-consisted field (SCF) cycle, the update of the orbitals and the update of the dressing parameters are performed jointly, so that the orbital shapes and the effective dipole displacement are optimized simultaneously and on equal footing \cite{Riso2022}.

For the conventional coherent-state QED-HF formulation, the SCF procedure follows the same overall structure. It requires no additional variational parameters beyond the orbital coefficients. At each iteration, the one-particle reduced density matrix $\mathbf{P}$, constructed from the current orbital coefficients as $P_{\mu\nu} = \sum_i C_{\mu i} C_{\nu i}^*$ (sum over occupied orbitals, where $\mu$ and $\nu$ label atomic basis functions), is used to update the displacement parameter $z$.

Since $z$ is fixed by the current density at each step rather than optimized through a separate stationarity condition, it is not an independent variational degree of freedom. The one-electron Hamiltonian is rebuilt at every iteration because it acquires a density-dependent dipole self-energy (DSE) contribution from the cavity \cite{Riso2022}. The two-electron part of the effective potential receives analogous DSE-mediated Coulomb and exchange corrections, and the augmented Fock matrix is diagonalized through the standard Roothaan--Hall generalized eigenvalue problem. The SCF cycle is considered converged once the change in total energy drops below the prescribed thresholds. As noted above, however, this simpler scheme introduces orbital origin dependence for charged systems: the Fock matrix eigenvalues depend on the choice of coordinate origin even though the total energy remains origin-invariant.

The SC-QED-HF approach offers several crucial advantages over both Fock-state QED-HF and coherent-state QED-HF. It restores full translational invariance: both energies and orbital coefficients are origin-independent, even for charged molecular species. This is verified numerically by displacing the entire molecular system through space and confirming energy and orbital invariance to machine precision~\cite{Riso2022} as well as serving as reliable reference states for subsequent correlated treatments such as cavity QED coupled-cluster theory~\cite{Haugland2020} or cavity QED configuration interaction~\cite{McTague_2022}.

Both QED-HF and SC-QED-HF generate cavity-modified potential energy surfaces (PESs) that are parametrized by the nuclear coordinates $\{\mathbf{R}_A\}$. These surfaces encode how the cavity field reshapes the energy landscape governing molecular structure, reactivity, and photophysics. By computing PESs as functions of nuclear geometry, scanning bond lengths, angles, and other internal coordinates while self-consistently solving the QED-HF or SC-QED-HF equations at each geometry, one can systematically identify cavity-induced phenomena: shifted equilibrium geometries, modified vibrational frequencies and force constants, and altered reaction barriers and transition-state structures.

Beyond the mean-field level, correlated methods such as cavity QED coupled-cluster theory~\cite{Haugland2020}, cavity QED configuration interaction~\cite{Vu_2024}, and cavity QED density functional theory~\cite{Aklilu_2026} systematically capture both electron-electron and electron-photon correlation effects. These methods are necessary for accurate predictions in systems with strong or multi-reference correlations, enabling consistent determination of excitation energies, oscillator strengths, and transition properties. Together, these approaches form a general platform for studying and designing molecular polaritons across the weak, strong, and ultra-strong coupling regimes, connecting quantum optics with chemistry and materials science.

Numerical implementations of QED-HF and SC-QED-HF methods in quantum chemistry packages have become available only recently, reflecting the emerging nature of {\it ab initio} cavity quantum electrodynamics (QED) simulations. At present, several open-source packages support quantum electrodynamics extensions of electronic structure methods. Notable examples include the \texttt{eT} program \cite{Folkestad_2020}, which implements QED-augmented self-consistent field approaches; \texttt{ExaChem}, a high-performance scalable framework with QED-coupled cluster capabilities \cite{Kowalski_2021}; \texttt{Octopus}, a computational framework for exploring light-driven phenomena including cavity QED effects \cite{TancogneDejean2020}; \texttt{hilbert} \cite{DePrince2020}, a plug-in to \texttt{Psi4} code \cite{Psi4}; and \texttt{OpenMS} \cite{OpenMS}, a multiscale solver developed at Los Alamos National Laboratory that couples Maxwell–Schr\"odinger dynamics with open quantum system techniques and provides interfaces to electronic structure packages such as \texttt{PySCF}~(Python-based Simulations of Chemistry Framework) \cite{PySCF}.
In this work, we employ \texttt{OpenMS} due to its seamless integration with \texttt{PySCF}, which serves as the foundation for our subsequent calculation of the diagonal Born–Oppenheimer correction.

\section{Evaluation of Diagonal Born-Oppenheimer Correction}
\label{dboc_theory}

In the Born-Oppenheimer (BO) approximation the nuclear coordinates
$\{\mathbf{R}_A\}$ are held fixed in the electronic Schr\"odinger
equation, and the resulting nonadiabatic couplings between electronic
and nuclear motion are neglected. While this simplification is often effective, it becomes quantitatively inadequate for systems containing light atoms such as hydrogen, where nuclear motion plays a more pronounced role. Achieving spectroscopic accuracy in such cases therefore requires explicit corrections beyond the BO framework.

The leading-order contribution is given by the diagonal Born-Oppenheimer correction (DBOC), which accounts for the expectation value of the nuclear kinetic energy operator with respect to the electronic wavefunction. In hydrogen-containing molecules, the DBOC is typically larger than both relativistic and quantum electrodynamics contributions, ans its inclusion is essential for accurate calculations of vibrational frequencies, dissociation energies, and thermochemical properties~\cite{Valeev_2003,Gauss_2006}.

Within the adiabatic approximation, the total molecular energy at fixed nuclear geometry $\mathbf{R}_{A}$ is $E_{\mathrm{ad}}(\mathbf{R}_{A})$. The DBOC modifies this to yield the effective potential energy surface $V(\mathbf{R}_{A})$ experienced by nuclei:
\begin{eqnarray}
V(\mathbf{R}_{A}) = E_{\mathrm{ad}}(\mathbf{R}_{A}) + E_{\mathrm{DBOC}}(\mathbf{R}_{A}),
\label{eq:total_PES}
\end{eqnarray}
where $E_{\mathrm{DBOC}}$ is defined as
\begin{eqnarray}
\label{eq:DBOC_definition}
E_{\mathrm{DBOC}}(\mathbf{R}_{A}) = \langle \Psi_e | \hat{T}_n | \Psi_e \rangle 
\\
\nonumber
= -\frac{1}{2} \sum_{A} \frac{1}{M_A} \left\langle \Psi_e \middle| \nabla^2_A \middle| \Psi_e \right\rangle.
\end{eqnarray}
Here $\hat{T}_n = -\frac{1}{2}\sum_A M_A^{-1} \nabla_A^2$ is the nuclear kinetic energy operator, $M_A$ is the mass of nucleus $A$ (in atomic units), $\nabla^2_A$ is the Laplacian with respect to nucleus coordinates, and $\Psi_e(\mathbf{r};\mathbf{R}_{A})$ is the electronic wavefunction parametrically dependent on nuclear positions $\mathbf{R}_{A}$. 

Physically, the DBOC accounts for the mutual influence of electronic and nuclear motion beyond the simple Born-Oppenheimer separation. As nuclei vibrate, the electronic wavefunction adiabatically adjusts to the changing potential; DBOC quantifies the energetic cost of this relaxation. In optical cavities, where strong light-matter coupling modifies electronic structure, cavity-dressed wavefunctions $\Psi_e^{\mathrm{cav}}$ from QED-HF or SC-QED-HF methods are expected to alter DBOC. This could impact vibrational %zero-point
energies, isotope effects, and reaction rates in polaritonic chemistry, motivating our extension of DBOC to cavity QED.

Direct evaluation of $\langle \Psi_e | \nabla^2_A | \Psi_e \rangle$ requires solving coupled-perturbed equations for analytical derivatives. This rigorous yet algebraically complex approach becomes intractable for cavity methods because of the orbital-dependent photonic parameters $\{\eta_{p\sigma}\}$. Instead, we adopt a finite-difference scheme using wavefunction overlaps at displaced geometries, which is generalizable to any electronic structure method and avoids explicit differentiation \cite{Valeev_2003}.

For nucleus $A$, we displace its position by $\pm\delta R$ along Cartesian direction $i \in \{x,y,z\}$, solving for $\Psi_e$ at each geometry. The second derivative of $\Psi_e$ is approximated via symmetric finite differences \cite{Valeev_2003}:
\begin{eqnarray}
\left\langle \Psi_e \middle| \frac{\partial^2}{\partial R_{A,i}^2} \middle| \Psi_e \right\rangle_{\!\mathbf{R}_0} 
\approx \frac{ S^{(+)}_{A,i} + S^{(-)}_{A,i} - 2 }{\delta R_{i}^2},
\label{eq:laplacian_overlap}
\end{eqnarray}
where $S^{(\pm)}_{A,i} = \langle \Psi_e(\mathbf{R}_A) | \Psi_e(\mathbf{R}_A \pm \delta \mathbf{R}_i) \rangle$ are wavefunction overlaps. Summing over directions yields the Laplacian:
\begin{eqnarray}
\label{grad}
\left\langle \Psi_e \middle| \nabla^2_A \middle| \Psi_e \right\rangle 
\approx \sum_{i=x,y,z} \frac{ S^{(+)}_{A,i} + S^{(-)}_{A,i} - 2 }{\delta R_{i}^2}.
\label{eq:laplacian_components}
\end{eqnarray}
The total DBOC then takes the form of Eq. (\ref{eq:DBOC_definition}). 
%\begin{eqnarray}
%\label{dboc}
%E_{\mathrm{DBOC}}= -\frac{1}{2} \sum_{A} \frac{1}{M_A} \left\langle \Psi_e \middle| \nabla^2_A \middle| \Psi_e \right\rangle.
%\label{eq:DBOC_final}
%\end{eqnarray}

In the Hartree-Fock approach for many electron wave-function $\Psi_e$ is a Slater determinant. The overlap $S^{(\pm)}_{A,i}$ factorizes into spin components:
\begin{eqnarray}
S^{(\pm)}_{A,i} = S_\alpha^{(\pm)} S_\beta^{(\pm)}, \quad 
S_\sigma^{(\pm)} = \det \left[ \mathbf{C}^{\sigma \dagger}_0 \mathbf{S}_{12}^{(\pm)} \mathbf{C}^\sigma_\pm \right],
\end{eqnarray}
where $\mathbf{C}^\sigma_0$ and $\mathbf{C}^\sigma_\pm$ are orbital coefficient matrices at $\mathbf{R}_0$ and displaced geometries, respectively, and $\mathbf{S}_{12}^{(\pm)}$ is the atomic orbital overlap matrix between these geometries. The matrix elements are available with the built-in methods of \texttt{PySCF}. For restricted Hartree-Fock (RHF) calculations $S_\alpha = S_\beta$, and consequently $S^{(\pm)}_{A,i} = (S_\alpha^{(\pm)})^2$. The \texttt{PySCF} package also enables the evaluation of the matrix element in Eq.~\eqref{eq:laplacian_overlap} not only for a single-determinant SCF wave function but also for a multi-determinant expansion obtained within the configuration interaction method including single and double excitations (CISD). This approach enables a consistent treatment of electron correlation effects in both energies and the required molecular properties.

The computation proceeds in two stages. First, the electronic wave function $\Psi_{e}(\mathbf{r};\mathbf{R}_{A})$ is obtained at the QED-HF or strong coupling SC-QED–HF level using the built-in solvers of the \texttt{OpenMS} package. Upon convergence of the self-consistent field (SCF) cycle, all one- and two-electron integrals in the chosen electronic basis set are available, along with the molecular orbital energies, $\bm{\varepsilon}$, and the matrix of molecular orbital coefficients $\bm{C}$. All of these quantities consistently incorporate cavity effects at the selected mean‑field level of theory.

In the second stage, these quantities are used to initialize a standard configuration interaction calculation including single and double excitations (CISD) as implemented in \texttt{PySCF} with SCF parameters obtained at the previous step. This treatment accounts for static and dynamic electron correlation in the energy and associated properties, while the electron–photon coupling remains described at the mean-field level. This approach is analogous to the special case of the quantum electrodynamics coupled-cluster hierarchy denoted as QED-CC($2,0$) in \cite{Haugland2020}. Within this approach, electronic excitations up to doubles are included, but photon excitations beyond the coherent-state reference of QED‑HF (or SC‑QED‑HF) are omitted. Consequently, electron–photon correlations are captured only implicitly through the polarized mean-field orbitals, while explicit many-body entanglement between electronic and photonic degrees of freedom is neglected.

A fully correlated treatment of electron–photon interactions would necessitate more advanced many‑body methods, such as cavity full configuration interaction (QED-FCI), QED many-body perturbation theory~\cite{ElMoutaoukal2025}, or electron-boson coupled-cluster approaches that explicitly include excitations in the photonic Fock space~\cite{PhysRevResearch.2.023262}. Incorporating such explicit electron-photon correlation effects into the evaluation of finite nuclear mass corrections lies beyond the scope of the present work. Accordingly, we restrict our analysis to the QED-CI(2,0) level of theory, which captures electronic correlation while retaining the photon field at the mean-field level.

Finally, for each nucleus $A$ and Cartesian direction $i$, we first compute $\Psi_e$ at $\mathbf{R}_A \pm \delta \mathbf{R}_i$ at the QED-CI(2,0) level of theory and then evaluate $S_{A,i}^{(\pm)}$ using the determinant formula above. The Laplacian components are constructed according Eq.~(\ref{eq:laplacian_components}), and $\Delta E_{\mathrm{DBOC}}$ is assembled from Eq.~(\ref{eq:DBOC_definition}).

\section{Single atoms in cavity}
\label{results_atoms}

For a single atom, the finite nuclear mass correction, also known as the QED recoil effect~\cite{Shabaev1985, PhysRevA.57.59}, is identical to the diagonal Born-Oppenheimer correction. This correction arises from the expectation value of the nuclear kinetic energy operator evaluated with the many-electron wave function. It has recently been computed for one- and two-electron systems, including H, He, H$^{-}$, and H$_2^{+}$, within a stochastic variational method framework~\cite{Ahrens_2021} based on explicitly correlated Gaussian basis functions~\cite{Nair_2025}.

In this work we employ configuration interaction with single and double excitations (CISD) built upon QED-HF or strong-coupling QED-HF reference wave functions within the coherent-state representation, as detailed in Section~\ref{theory}. This approaches, denoted as QED-CI(2,0)/QED-HF and QED-CI(2,0)/SC-QED-HF, respectively, are then applied for $^4$He, $^1$H$^{-}$, and $^9$Be atoms, incorporates electronic correlation at the CISD level while treating photon degrees of freedom at the mean-field level. The notation QED‑CI(2,0) explicitly indicates the absence of direct electron‑photon correlation terms in the wave‑function ansatz, as no photon‑number‑changing excitations are included in the configuration space.

The light-matter coupling strength $\lambda$ is varied over the interval $0.0 \le \lambda \le 0.1$~a.u., following Ref.~\cite{Nair_2025}. Values of $\lambda \lesssim 0.05$~a.u. correspond to the strong-coupling regime, while $\lambda \sim 0.1$~a.u. approaches the ultrastrong-coupling threshold; both regimes are experimentally accessible in state-of-the-art plasmonic and photonic cavity systems and are widely used as reference points in {\it ab initio} cavity QED studies.

\begin{figure}[h!]
    \centering
    \includegraphics[width=0.8\linewidth]{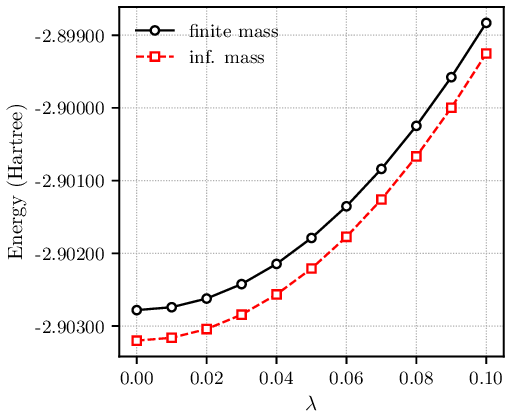}
    \caption{Mass dependence of the 
$^4$He energy as a function of the light–matter coupling strength $\lambda$.}
    \label{fig:1}
\end{figure}

\begin{figure}[h!]
    \centering
    \includegraphics[width=0.8\linewidth]{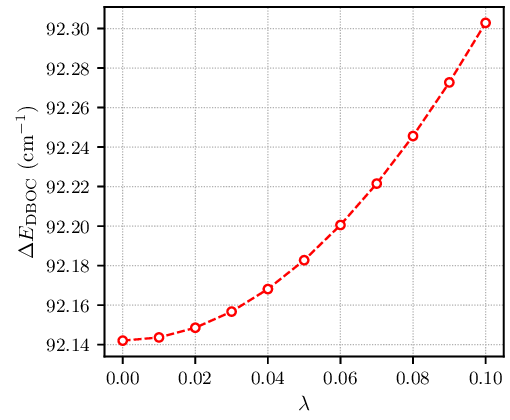}
    \caption{DBOC for 
$^4$He as a function of the light–matter coupling strength $\lambda$.}
    \label{fig:2}
\end{figure}

Figure~\ref{fig:1} shows the dependence of the $1^1S$ ground-state energy of the helium atom on the light-matter coupling strength $\lambda$. The two depicted curves correspond to the cases of infinite and finite nuclear masses. All calculations for atomic systems were performed using the correlation-consistent Dunning aug-cc-pV5Z basis sets. The curve including the finite nuclear mass correction (DBOC) is shown as a solid line in Fig.~\ref{fig:1} and is clearly shifted relative to the infinite‑nuclear‑mass case.
The finite nuclear mass correction (DBOC) itself is shown in more detail in Fig.~\ref{fig:2}.

It is important to note that, at zero light--matter coupling ($\lambda = 0.0$), the finite nuclear mass correction evaluated numerically using Eq.~(\ref{grad}) at the QED-CI(2,0)/QED-HF level equals $0.00041950$ a.u. and agrees closely with the well-established variational benchmark value of $0.00041982$ a.u. for $^{4}$He~\cite{Drake_book}. As the coupling strength increases, the resulting dependence, both for the total energy and for this correction, is in full agreement with the results reported in \cite{Nair_2025}~(see Fig. 3 therein). 

\begin{figure}[h!]
    \centering   \includegraphics[width=0.8\linewidth]{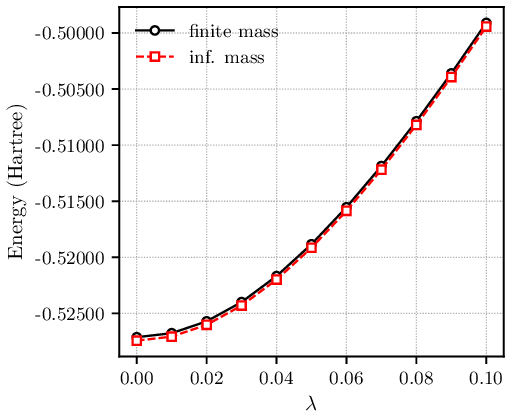}
    \caption{Mass dependence of the 
$^1$H$^{-}$ energy as a function of the light–matter coupling strength $\lambda$.}    \label{fig:3}
\end{figure}

\begin{figure}[h!]
    \centering
    \includegraphics[width=0.8\linewidth]{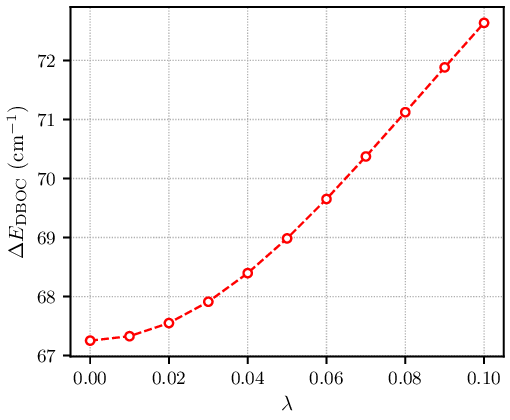}
    \caption{DBOC for 
$^1$H$^{-}$ as a function of the light–matter coupling strength $\lambda$.}
    \label{fig:4}
\end{figure}

The next two‑electron system we consider is the hydrogen anion, H$^{-}$. The results for the energy dependence (which includes the infinite and finite nuclear mass cases) and for the DBOC itself are shown in Figs.~\ref{fig:3} and \ref{fig:4}, respectively.

Upon detailed comparison of the calculated values with the results of Ref.~\cite{Nair_2025}, a slight deviation is observed starting at coupling strengths $\lambda \ge 0.05$. This discrepancy arises because in Ref.~\cite{Nair_2025} the finite-mass correction is treated non-perturbatively and electron–photon correlations are included beyond the mean-field level within a variational framework. In contrast, our approach incorporates finite-mass effects perturbatively, in close analogy to standard quantum-chemical treatments \cite{Handy_1986}. Consequently, our results indicate that, in contrast to helium, a non-perturbative treatment of finite-mass effects is more appropriate for lighter systems such as $\mathrm{H}^-$ in the strong-coupling regime.

\begin{figure}[h!]
    \centering
    \includegraphics[width=0.8\linewidth]{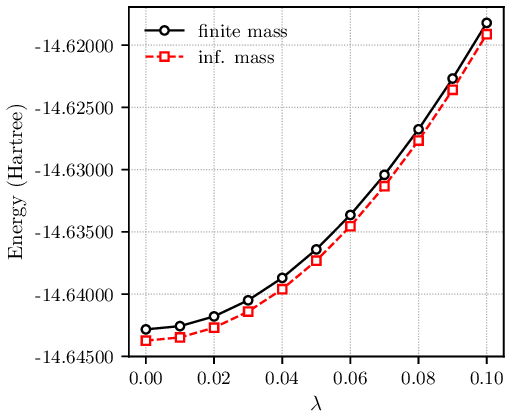}
        \caption{Mass dependence of the 
$^9$Be energy as a function of the light–matter coupling strength $\lambda$.}
    \label{fig:5}
\end{figure}

\begin{figure}[h!]
    \centering
    \includegraphics[width=0.8\linewidth]{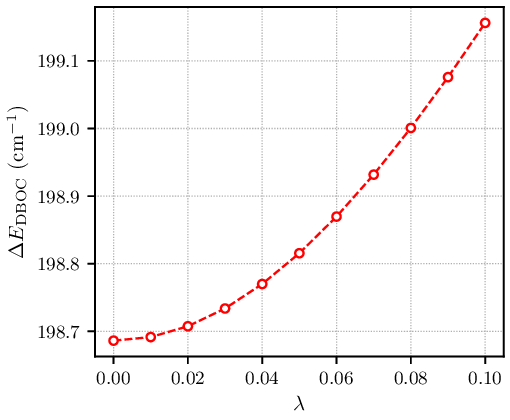}
    \caption{DBOC for 
$^9$Be as a function of the light–matter coupling strength $\lambda$.}
    \label{fig:6}
\end{figure}

For the four-electron $^{9}$Be atom, analysis reveals a dependence of both the ground-state energy and the finite nuclear mass correction on the light-matter coupling strength $\lambda$ that is similar to that observed for the previous systems, see Figs.~\ref{fig:5} and \ref{fig:6}.

At zero coupling, $\lambda = 0.0$, the computed DBOC is consistent with the established reference value for beryllium \cite{Stanke2009}. As $\lambda$ increases within the considered range, both the total energy and the correction exhibit a regular and nearly linear behavior. In contrast to lighter systems such as H$^{-}$, the perturbative treatment of finite nuclear mass effects in $^{9}$Be closely resembles the behavior observed for helium, remaining well behaved across the considered range of light–matter coupling strengths. For all atoms studied, the recoil effect grows with increasing coupling strength. Varying $\lambda$ from zero to $0.1$ changes the correction for heavier atoms at the level of tenths of an inverse centimeter, while for the lightest system considered the variation is on the order of five inverse centimeters.

%Нужно ли таблицы со значениями - одну: строки $\lambda$, столбцы атомы? было бы удобно
% было написано о линейной зависимости для аниона водорода, но по графикам экспонента или парабола. Можно ли построить заодно экстраполяцию для DBOC на графиках, где DBOC отдельно?  было бы здорово если для всех одинаково (такова природа). $\lambda$ - безразмерный, значит коэффициент перед тоже - отношение масс электрона и ядра? Перед "экспонентой" сдвиг энергии при $\lambda=0$

\section{Diatomic molecules}
\label{results_molecules}

In this section we investigate finite nuclear mass effects for three two-atomic hydrides H$_{2}$, LiH and HF within the QED-CI(2,0) approach. The ano-PVTZ basis set (Atomic Natural Orbitals) was chosen because it provides a good balance between accuracy and computational cost for diatomic molecules. Compared to conventional PVTZ basis sets, ano-PVTZ is more flexible and systematically optimized using atomic natural orbitals, which offer a more compact and physically meaningful representation of the electronic structure \cite{Neese_2010}. This yields both faster convergence and enhanced accuracy in energies and molecular properties.

\begin{figure}[h!]
\centering
\includegraphics[width=0.9\linewidth]{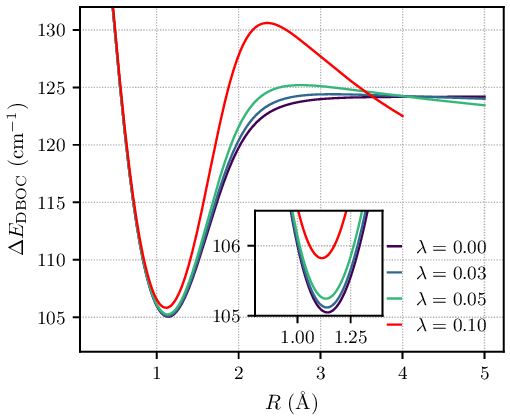}
\begin{center}
\text{(a)}
\end{center}
\includegraphics[width=0.9\linewidth]{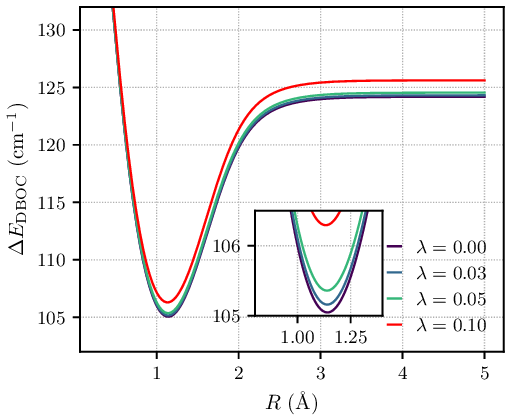}
\begin{center}
\text{(b)}
\end{center}
\caption{DBOC as a function of internuclear distance $R$ for H$_2$ molecule in a single-mode cavity and cavity field polarized along (a) and perpendicular (b) to the molecular axis. Results obtained at the QED-CI(2,0)/QED-HF level of theory. The insets in the figures reveal the finer details of the curve behavior in the vicinity of the minimum for various $R$ values.}
\label{fig:7}
\end{figure}

\begin{figure}[h!]
\centering
\includegraphics[width=0.9\linewidth]{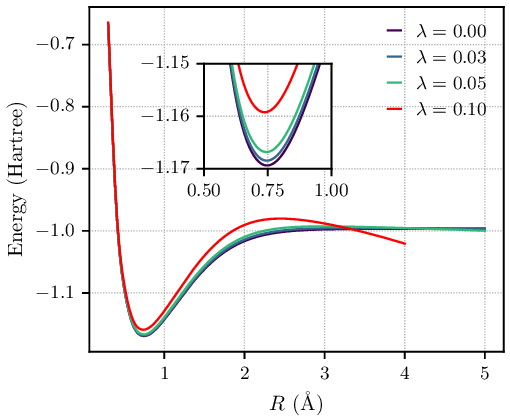}
\begin{center}
\text{(a)}
\end{center}
\includegraphics[width=0.9\linewidth]{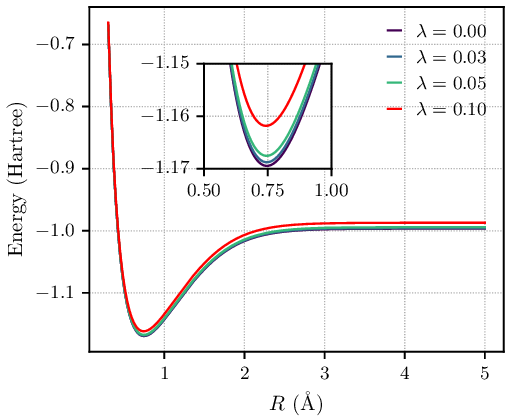}
\begin{center}
\text{(b)}
\end{center}
\caption{Potential energy curves for the H$_2$ molecule coupled to a single-mode cavity, with the cavity field polarized along (a) and perpendicular (b) to the molecular axis. Results obtained at the QED-CI(2,0)/QED-HF level of theory. The insets in the figures reveal the finer details of the curve behavior in the vicinity of the minimum for various $R$ values.}
\label{fig:8}
\end{figure}

The first molecule considered is the hydrogen molecule. The diagonal Born–Oppenheimer correction (DBOC), calculated as a function of internuclear distance and light–matter coupling strength for polarizations parallel and perpendicular to the molecular axis, is shown in Figs.~\ref{fig:7}(a) and \ref{fig:7}(b), respectively.
The corresponding potential energy curves are presented in Figs.~\ref{fig:8}~(a), \ref{fig:8}~(b).

Using a self‑consistent QED‑HF reference in the coherent‑state representation for all calculations shown in Figs.~\ref{fig:7} and~\ref{fig:8}, we find that, for the parallel field configuration, both the potential energy curves and the DBOC approach an incorrect asymptotic limit in the dissociation regime with increasing coupling strength. In contrast, no such behavior is observed for the perpendicular orientation. In this case, the asymptote of the potential energy curve correctly approaches the sum of the energies of two isolated atoms in the cavity, while the DBOC tends to the sum of the finite-mass corrections of the individual atoms.

For the parallel orientation, this behavior is related to the sensitivity of QED-HF to the choice of the origin and is analogous to the well-known behavior of diatomic molecular potentials in strong magnetic fields when finite basis sets are employed \cite{Stopkowicz2015,zalialiutdinov2025exploring}. In particular, it has been shown that for a perpendicular orientation of a strong magnetic field, potential energy curves also approach an incorrect dissociation limit, and this problem is resolved only in the limit of an infinite basis set \cite{zalialiutdinov2025exploring}. In practical finite-basis calculations, atomic orbitals are therefore replaced by so-called London orbitals or gauge-including atomic orbitals (GIAOs), which are defined as the product of an atomic orbital (e.g., a Gaussian function) and a phase factor depending on the magnetic field strength \cite{London1937}. Such calculations have become feasible in quantum chemistry only relatively recently and allow one to restore gauge invariance in finite-dimensional basis sets \cite{Irons2017,Monzel2022}.

\begin{figure}[h!]
\centering 
\includegraphics[width=0.9\linewidth]{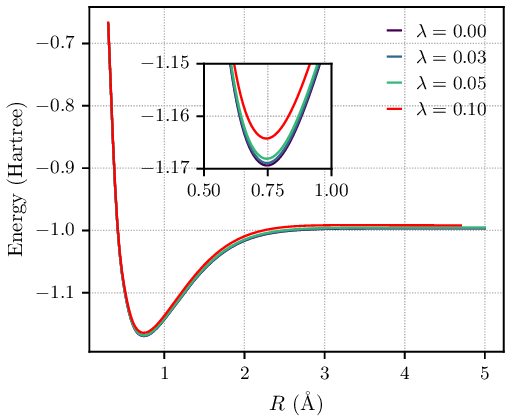}
\begin{center}
\text{(a)}
\end{center}
\includegraphics[width=0.9\linewidth]{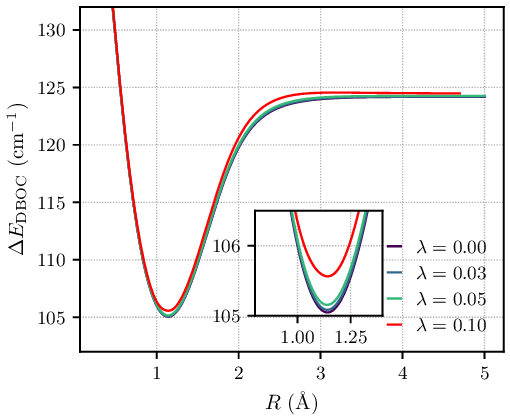}
\begin{center}
\text{(b)}
\end{center}
\caption{PEC (a) and the corresponding DBOC (b) as a function of internuclear distance $R$ for H$_2$ molecule in a single-mode cavity and field polarized along the molecular axis. Results obtained at the QED-CI(2,0)/SC-QED-HF level of theory. The insets in the figures reveal the finer details of the curve behavior in the vicinity of the minimum for various $R$ values.}
\label{fig:12}
\end{figure}

As can be seen from Figs.~\ref{fig:7}-\ref{fig:8}, although in the coherent-state representation there is no dependence on the number of photons in the basis, in the strong-coupling regime problems remain because the transformation given by Eq.~(\ref{CS1})~(which is essentially analogous to the introduction of London orbitals) is applied globally to all electronic orbitals. Within the SC-QED-HF framework, this transformation acts on each orbital individually, exactly as in the GIAO approach. Subsequent CI calculations built on top of the SC-QED-HF reference resolve the problem for strong coupling in the parallel configuration, as demonstrated in Fig.~\ref{fig:12}. Notably, at the equilibrium geometry $R_\text{eq}$, both CI approaches,
employing either a QED-HF or an SC-QED-HF reference, yield very similar results.

The maximum difference in the diagonal Born–Oppenheimer correction between the zero-coupling and $\lambda = 0.1$ cases occurs away from $R_\text{eq}$, at $R \approx 2.02$,\AA, and amounts to $0.9$ cm$^{-1}$, see Fig.~\ref{fig:12}~(b). The fact that this maximum is located well past the equilibrium geometry indicates that the photon field preferentially modifies the DBOC in the region of stretched bond lengths, where the electronic wave function is more diffuse and thus more susceptible to cavity-induced perturbations. Nevertheless, the overall magnitude of the effect remains small across the entire potential energy curve, confirming that at $\lambda = 0.1$ the cavity coupling acts as a weak perturbation to the adiabatic nuclear dynamics of H$_2$.

\begin{figure}[h!]
\centering 
\includegraphics[width=0.9\linewidth]{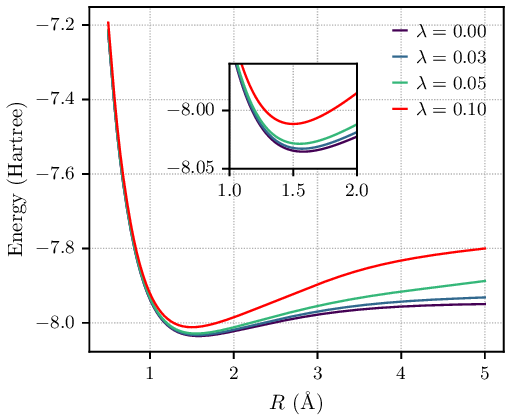}
\begin{center}
\text{(a)}
\end{center}
\includegraphics[width=0.9\linewidth]{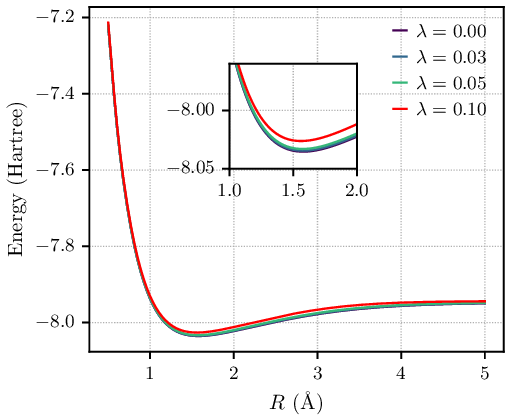}
\begin{center}
\text{(b)}
\end{center}
\caption{Potential energy curves for the LiH molecule coupled to a single-mode cavity with field polarized along (a) and perpendicular (b) to the molecular axis. Results obtained at the QED-CI(2,0)/QED-HF level of theory.}
\label{fig:13}
\end{figure}

\begin{figure}[h!]
\centering 
\includegraphics[width=0.9\linewidth]{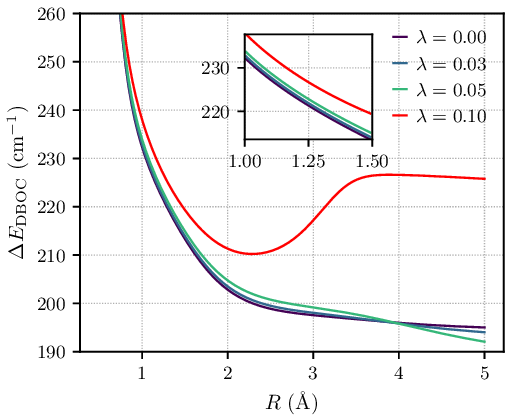}
\begin{center}
\text{(a)}
\end{center}
\includegraphics[width=0.9\linewidth]{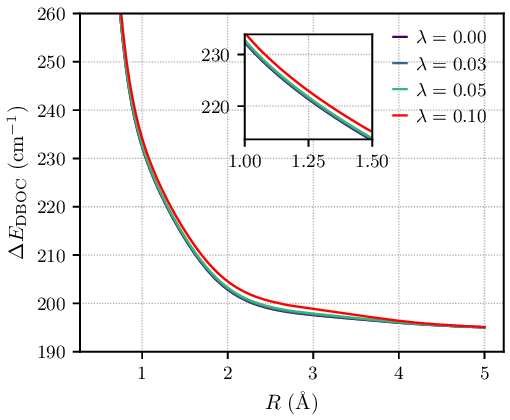}
\begin{center}
\text{(b)}
\end{center}
    \caption{DBOC as a function of internuclear distance $R$ for LiH molecule in a single-mode cavity and  field polarized along (a) and perpendicular (b) to the molecular axis. Results obtained at the QED-CI(2,0)/QED-HF level of theory.}
    \label{fig:14}
\end{figure}

For the lithium hydride molecule, the resulting potential energy curves for the parallel and perpendicular field configurations are shown in Figs.~\ref{fig:13} (a) and (b), respectively. 
The corresponding DBOC profiles as functions of the internuclear distance for different values of the coupling strength $\lambda$ are presented in Fig. \ref{fig:14}. As in the case of the hydrogen molecule, starting from $\lambda \approx 0.05$, the use of a QED-HF reference leads to an incorrect behavior in the dissociation limit for the parallel configuration.

The maximum difference in the DBOC between the zero-coupling limit and the strongest cavity coupling considered here remains on the order of $1$ cm$^{-1}$, see Fig.~\ref{fig:14} (b). While this is a small correction in absolute terms, it is worth emphasizing that for H$_2$ and LiH high-precision experimental benchmarks exist against which theoretical predictions can be tested well below the cm$^{-1}$ level. The dissociation energy of H$_2$, $D_0(\text{H}_2) = 35999.582894(25)$ cm$^{-1}$~\cite{Cheng2018}, has been measured with an absolute uncertainty of $2.5 \times 10^{-5}$ cm$^{-1}$, and for LiH an extensive compilation of precision spectroscopic data provides vibrational energy levels accurate to the sub-cm$^{-1}$ level~\cite{Stwalley1993}. The cavity‑induced modification of the DBOC, though small on the scale of typical quantum‑chemical errors, turns out to be very significant in a regime that is, in principle, experimentally resolvable for these benchmark systems.

\begin{figure}[h!]
\centering 
\includegraphics[width=0.9\linewidth]{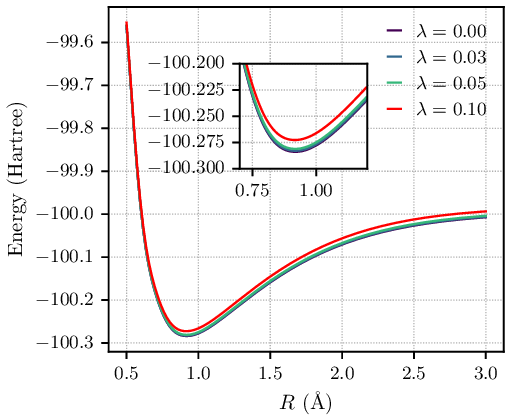}
\begin{center}
\text{(a)}
\end{center}
\includegraphics[width=0.9\linewidth]{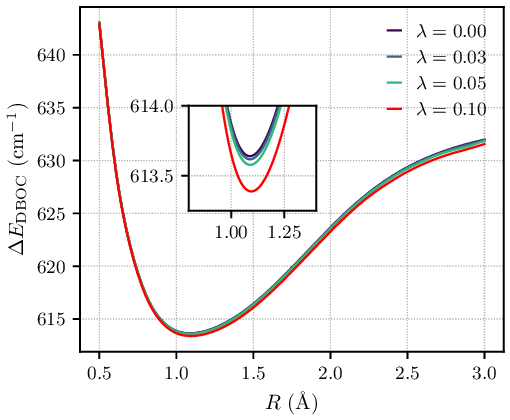}
\begin{center}
\text{(b)}
\end{center}
    \caption{PECs (a) and DBOCs (b) for the HF molecule coupled to a single-mode cavity with the cavity field polarized perpendicular to the molecular axis. Results obtained at the QED-CI(2,0)/QED-HF level of theory.}
    \label{fig:17}
\end{figure}

We also performed calculations for the hydrogen‑fluoride (HF) molecule, motivated by recent cavity‑QED studies of its spectroscopic and dynamical behavior \cite{Schnappinger2023vibropolaritonic,Lindoy2025dissociation}. In particular, Fig.~\ref{fig:17}~(a) shows the computed potential energy curves, while Fig.~\ref{fig:17}~(b) presents the corresponding diagonal Born–Oppenheimer corrections for several light–matter coupling strengths.

For the HF molecule, the situation is similar to the previous cases, but the absolute effect is smaller. The maximum difference in the DBOC between the cavity-free and $\lambda = 0.1$ cases amounts to only $\approx 0.25$ cm$^{-1}$. This reduction is physically transparent. The diagonal Born-Oppenheimer correction scales as $M_{A}^{-1}$, so replacing hydrogen by the heavier fluorine nucleus suppresses the nuclear kinetic energy contribution and, consequently, the sensitivity of the DBOC to the cavity field. Nevertheless, even this smaller shift is not beyond experimental error. The rovibrational spectrum of HF has been characterized with sub-0.01 cm$^{-1}$ accuracy by direct potential fits based on tens of thousands of spectroscopic line positions~\cite{Coxon2015}. Thus, cavity-induced modifications of the DBOC at the $0.1$ cm$^{-1}$ level are in principle spectroscopically resolvable.

\section{Polyatomic molecules}
\label{poly}

In this section, we extend our analysis to polyatomic molecules by considering two representative systems - ammonia (NH$_3$) and formaldehyde (CH$_2$O). These compounds were chosen for their distinct cavity-QED-relevant properties: a tunneling inversion mode in the THz regime~\cite{Harde2001} and a strongly polar carbonyl stretching mode in the mid-infrared range~\cite{Mordovina2025hessian}.

The ammonia molecule has a significant dipole moment ($\approx 1.47$~D), making it highly sensitive to the electric field of a cavity. It exhibits an ``umbrella'' inversion mode, corresponding to tunneling between two equivalent configurations through the plane of NH$_3$. The energy of this tunneling transition lies in the millimeter-to-infrared range, overlapping with typical microwave and THz cavity frequencies. These properties make NH$_3$ an ideal candidate for investigating strong and ultra-strong
light-matter coupling in cavity QED.

In turn, the formaldehyde molecule possesses a substantial permanent dipole moment ($\approx 2.33$~D), making it considerably more sensitive to the vacuum electric field of a cavity than ammonia. Furthermore, CH$_2$O serves as the simplest carbonyl-containing molecule, making it a prototypical system for studying cavity-induced modifications of chemical reactivity at a C$=$O bond~\cite{Schnappinger2025polarizability, Montillo2025hessian}. These properties render formaldehyde a compelling benchmark for probing vibrational strong coupling and cavity-mediated control of bond dynamics within the cavity QED framework.

The DBOC and one-dimensional potential energy curve for NH$_3$ were computed as follows. For each value of the light-matter coupling strength $\bm{\lambda}$ (directed along the molecular $C_3$ symmetry axis), the molecular geometry was fully optimized at the QED-HF level of theory until the forces on all atoms were driven to zero. Then the finite-difference DBOC calculation was performed at the equilibrium geometry. These calculations are carried out at the QED-CI(2,0) level of theory with QED-HF reference, consistent with the treatment used for atoms and the diatomic systems in previous sections, ensuring that the cavity-induced modifications of the DBOC are evaluated on equal footing. Geometry optimizations were performed using a fully numerical scheme for force gradients at the mean-field level. All calculations were carried out with the ano-PVTZ basis sets. At the equilibrium geometry, the cavity-induced shift in the DBOC amounts to $0.32$ cm$^{-1}$ for $\lambda = 0.1$ and $0.073$ cm$^{-1}$ for $\lambda = 0.5$, both relative to the field-free limit ($\lambda = 0$).

Of particular interest is the behavior of the Born–Oppenheimer diagonal correction for the umbrella inversion tunneling transition in ammonia. Inversion transition serves as a highly sensitive probe of the shape and height of the potential barrier, as well as of subtle non-adiabatic and cavity-induced effects on the nuclear dynamics. This fundamental process corresponds to the quantum tunneling of the nitrogen atom through the plane of the three hydrogen atoms, inter-converting two equivalent pyramidal $C_{3v}$ minima via a planar $D_{3h}$ transition state. 
Due to tunneling, the ground vibrational state is split into a symmetric and an antisymmetric component, giving rise to the well-known inversion doubling. The corresponding transition frequency lies in the microwave region at $23.787$\,GHz ($0.793$\,cm$^{-1}$), and has been determined with sub-kilohertz precision or ${\sim}3 \times 10^{-8}$\,cm$^{-1}$~\cite{Kukolich1967}.

\begin{figure}
    \centering
    \includegraphics[width=0.9\linewidth]{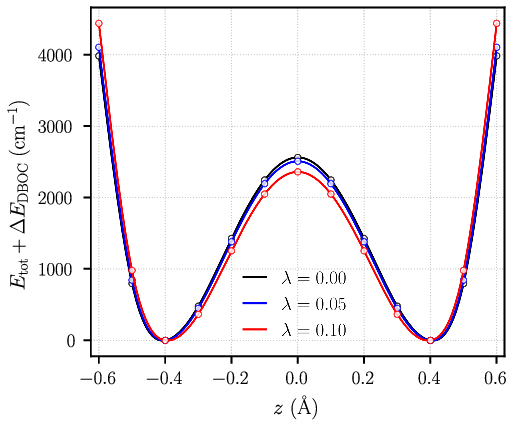}
\caption{One-dimensional potential energy curve for the umbrella inversion mode of NH$_3$, constructed by displacing the nitrogen atom along the $z$-axis with the N--H bond length fixed at $1.012~\text{\AA}$. The potential energy includes the QED-CI(2,0)/QED-HF electronic contribution and the diagonal Born-Oppenheimer correction.}
    \label{fig:umbrella_pec}
\end{figure}

\begin{figure}
    \centering
    \includegraphics[width=0.9\linewidth]{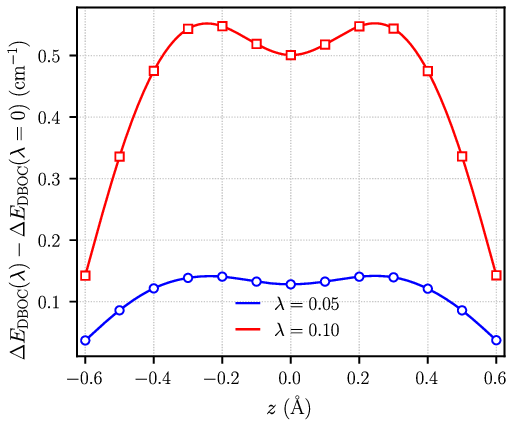}
\caption{Absolute change in DBOC for the 
NH$_3$ umbrella-inversion coordinate induced by light-matter coupling, relative to the 
cavity-free limit ($\lambda = 0$). The inversion coordinate is defined by displacing 
the nitrogen atom along the $z$ axis while keeping the N--H bond length fixed at 
$1.012$~\AA.}
\label{fig:dboc_shift}
\end{figure}

Figure \ref{fig:umbrella_pec} shows the one-dimensional potential energy curve for the umbrella inversion of NH$_3$, obtained by displacing the nitrogen atom along the $z$-axis while keeping the N–H bond length fixed at $1.012$ \AA. The total energy includes the QED-CI(2,0)/QED-HF electronic energy and the DBOC. The absolute cavity-induced change in the DBOC relative to the cavity-free case ($\lambda=0$), $\Delta E_{\mathrm{DBOC}}(\lambda)-\Delta E_{\mathrm{DBOC}}(\lambda=0)$, is presented in Fig.~\ref{fig:dboc_shift}. Within the considered displacement interval, the DBOC varies by a fraction of a cm$^{-1}$, representing a few to several tens of percent of the ammonia inversion transition frequency.

\begin{figure}
    \centering
    \includegraphics[width=0.9\linewidth]{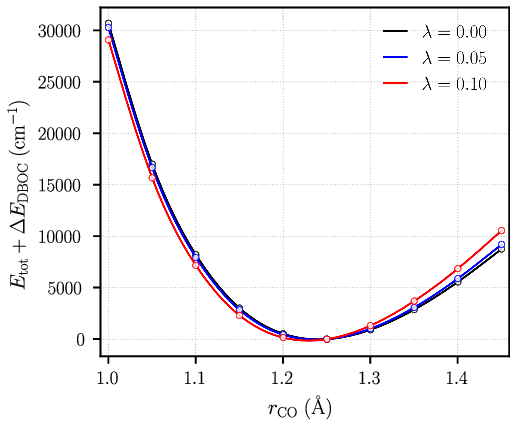}
\caption{One-dimensional potential energy curve for the C=O carbonyl
         stretching mode, obtained by displacing the carbon atom along
         the CO axis. The total energy includes the QED-CI(2,0)/QED-HF
         electronic contribution and the diagonal Born-Oppenheimer
         correction.}
    \label{fig:ch2o_pec}
\end{figure}

\begin{figure}
    \centering
    \includegraphics[width=0.9\linewidth]{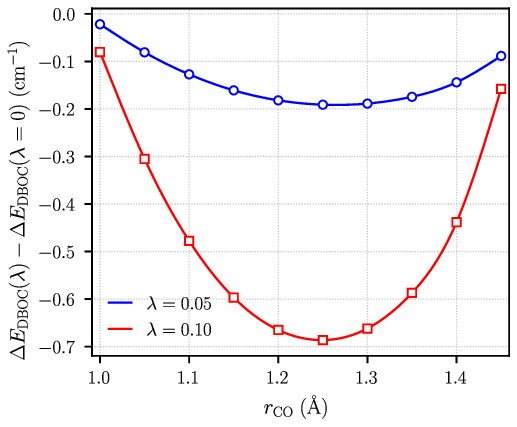}
\caption{Absolute change in the DBOC for the C=O carbonyl stretching
         mode induced by light-matter coupling, relative to the
         cavity-free limit ($\lambda = 0$).}
    \label{fig:ch20dboc_pec}
\end{figure}

The one-dimensional potential energy curve for the C=O carbonyl stretching mode of the CH$_2$O molecule is shown as a function of $r_{\rm CO}$ (the displacement of the carbon atom along the CO axis) in Fig.~\ref{fig:ch2o_pec}. The corresponding DBOC contribution is presented in Fig.~\ref{fig:ch20dboc_pec}.

For the equilibrium geometry of CH$_{2}$O, the cavity-induced shift in the DBOC amounts to $0.018$ cm$^{-1}$ for $\lambda = 0.1$, with the cavity mode polarized along the C=O bond axis. As illustrated in Figs.~\ref{fig:ch2o_pec} and~\ref{fig:ch20dboc_pec}, light-matter coupling induces a weakly coordinate-dependent modification of the potential energy curve as well as the DBOC along the C=O stretching coordinate, on the scale of a fraction of an inverse centimeter. Although the absolute cavity-induced shift of the DBOC is small, it is systematically non-zero across the entire C=O stretching coordinate and grows with the coupling strength~$\lambda$. This behavior is qualitatively consistent with the results obtained for the umbrella mode of NH$_3$, where the cavity-induced DBOC shift was found to be of the same order of magnitude (compare Figs.~\ref{fig:dboc_shift} and~\ref{fig:ch20dboc_pec}).     While the effect was studied at the single‑molecule level here, it is expected to be significant in the collective strong‑coupling regime (where many molecules interact coherently with a single cavity mode) that is relevant to current cavity chemistry experiments \cite{Ruggenthaler2023}.

\section{Conclusions}
\label{conclusion}

In this work, we develop and implement a general finite-nuclear-mass correction scheme for molecules embedded in cavities. The approach combines cavity quantum electrodynamics Hartree-Fock references, both QED-HF and SC-QED-HF, with a configuration-interaction treatment including single and double excitations, QED-CI(2,0). Within these approaches, the diagonal Born–Oppenheimer correction is evaluated numerically using wave‑function overlaps at displaced nuclear geometries. By construction, the scheme requires no analytic derivatives of photonic or orbital-dressing parameters and is therefore directly applicable to arbitrary electronic-structure methods within the cavity QED framework.

For atomic systems (He, H$^{-}$, Be), the cavity–induced modification of the finite–mass correction is found to be on the order of a few cm$^{-1}$. As a baseline, calculations were performed at zero coupling; our results reproduce field‑free benchmarks and remain consistent with existing non‑perturbative data within their expected regime of validity. The comparison indicates that, whereas a perturbative DBOC treatment on top of QED-CI(2,0) is quantitatively reliable for helium and beryllium, light one–electron systems such as H$^{-}$ in the strong–coupling regime, $\lambda \ge 0.05$, may require a genuinely non-perturbative description of nuclear–mass and electron–photon correlation effects. Moving to helium, the lightest two-electron atom, the results obtained with the present approach are in excellent agreement with those reported in Ref.~\cite{Nair_2025}, which confirms that the approximations introduced here remain well justified beyond the two-electron case.

For diatomic hydrides (H$_2$, LiH, HF), we have shown that the cavity vacuum reshapes both the potential–energy surfaces and the associated DBOC profiles, leading to shifts in dissociation limits and equilibrium vibrational properties on the scale of few cm$^{-1}$. We also found that coherent–state QED-HF references display unphysical behavior for fields aligned with the molecular axis in the strong–coupling regime. Application of the SC–QED–HF–based QED-CI(2,0) method restores correct asymptotics and translational invariance of both energies and orbitals. 

Our analysis of the polyatomic systems NH$_3$ and CH$_2$O further highlights that cavity-modified DBOC contributions can be significant in high-precision spectroscopic experiments. The DBOC is critically important for the ammonia tunneling frequency, as the contribution of finite nuclear masses and their motion can amount to several tens of percent depending on the light–matter coupling strength. Notably, the cavity-induced changes to the DBOC are comparable in magnitude to conventional quantum-electrodynamics corrections such as self-energy (SE) and vacuum-polarization (VP) shifts. This finding underscores that finite-mass effects have to be treated on equal footing with SE and VP corrections when targeting high-accuracy spectroscopy within the cavity QED framework~\cite{PhysRevA.63.024502,Sunaga2022,Janke2025}.

The QED-CI(2,0) framework introduced here provides the first systematic treatment of finite-nuclear-mass corrections in cavity QED, establishing a rigorous foundation for high-accuracy molecular spectroscopy in optical cavities. In the present implementation, the photon field is treated at the mean-field level. Extending the methodology to fully correlated electron-boson wave functions, including cavity FCI, QED-MBPT, and electron-boson coupled-cluster approaches, as well as to multi-mode and lossy cavities, represents a natural and straightforward continuation of this work. Such extensions will enable quantitatively predictive
descriptions of isotope effects, reaction rates, and tunneling phenomena in cavity-modified chemistry under realistic experimental conditions.

\section*{Acknowledgements}
This work was supported by the Russian Science Foundation under grant \textnumero{25-22-00643}.

\section*{Data Availability}
The data that support the findings of this study are available 
from the corresponding author upon reasonable request.

\bibliography{mybibfile}

\end{document}